\documentclass[aps,prl,twocolumn,superscriptaddress,preprintnumbers,amsmath,amssymb,showpacs]{revtex4}
\usepackage{graphicx}% Include figure files
\usepackage{dcolumn}% Align table columns on decimal point
\usepackage{bm}% bold math

\newcommand{\ba}{BaFe$_{2-2x}$Co$_{2x}$As$_2$}
\newcommand{\bao}{BaFe$_{1.84}$Co$_{0.16}$As$_2$}
\begin{document}

\title{Observation of a coherence peak and pair-breaking effects in THz conductivity of BaFe$_{2-2x}$Co$_{2x}$As$_2$}% Force line breaks with \\ 

\author{R. Vald\'{e}s Aguilar}
\email{rvaldes@pha.jhu.edu}
 \affiliation{The Institute of Quantum Matter, Department of Physics and Astronomy, The Johns Hopkins University, Baltimore, Maryland 21218}
 \author{L.S. Bilbro}
 \affiliation{The Institute of Quantum Matter, Department of Physics and Astronomy, The Johns Hopkins University, Baltimore, Maryland 21218}
 \author{S. Lee}
\affiliation{Department of Material Science and Engineering, University of Wisconsin, Madison, WI 53706}
\author{J. Jiang}
\affiliation{National High Magnetic Field Laboratory, Florida State University, Talahassee, FL 32310}
\author{J.D. Weiss}
\affiliation{National High Magnetic Field Laboratory, Florida State University, Talahassee, FL 32310}
\author{E.E. Hellstrom}
\affiliation{National High Magnetic Field Laboratory, Florida State University, Talahassee, FL 32310}
\author{D.C. Larbalestier}
\affiliation{National High Magnetic Field Laboratory, Florida State University, Talahassee, FL 32310}
\author{C.B. Eom}
\affiliation{Department of Material Science and Engineering, University of Wisconsin, Madison, WI 53706}
\author{N. P. Armitage}
\email{npa@pha.jhu.edu}
 \affiliation{The Institute of Quantum Matter, Department of Physics and Astronomy, The Johns Hopkins University, Baltimore, Maryland 21218}
 
 \date{\today}

\begin{abstract}
We report a study of high quality pnictide superconductor \bao\ thin films using time-domain THz spectroscopy.  Near T$_c$ we find evidence for a coherence peak and qualitative agreement with the weak-coupling Mattis-Bardeen form of the conductivity. At low temperature, we find that the real part of the THz conductivity is not fully suppressed and  $\sigma_2$ is significantly smaller than the Matthis-Bardeen expectation. The temperature dependence of the penetration depth $\lambda$ follows a power law with an unusually high exponent  of 3.1. We interpret these results as consistent with impurity scattering induced pair-breaking.   Taken together our results are strong  evidence for an extended s$\pm$ symmetry order parameter.
\end{abstract}

\pacs{74.25.Gz,74.70.Xa,74.25.-q}
\maketitle

The discovery of iron based superconductors \cite{kamihara2008iron,Rotter:2008ys,Fong-ChiHsu09232008,Wang2008538,Sefat:2008zr} has further increased interest in the physics of high temperature superconductivity. Among other things, this has stemmed from similarities in the phase diagram between high-T$_c$ cuprates and these new superconductors.  Both share a proximity and interplay with a magnetically ordered state, a superconducting dome as a function of doping and anomalous \textit{normal} state transport properties \cite{Yi2009uq,Chu2010kx}. However, significant differences also exist.   For instance, for the iron based materials the undoped parent compounds are metallic, while the cuprates are Mott insulators. Also, in the cuprates, a single band model has been considered a good starting point to describe the basic physical picture, whereas the iron pnictides have multiple occupied \textit{d} orbitals which make even a simple tight binding model more involved \cite{Cvetkovic09}.

Based on LDA first principles calculations \cite{Mazin:2008fk}, it was suggested that a natural symmetry for the order parameter was one in which the sign of the wave function was opposite for different sheets of the Fermi surface (FS), so called extended s-wave or s$\pm$. This was based on LDA-predicted band structure that have 2 almost circular electron and hole pockets separated by the wavevector $(\pi,\pi)$, which matches the magnetic ordering wavevector in the undoped materials. This Fermi surface geometry lends itself to a pairing scenario based on spin-wave exchange with that wavevector. Further analysis based on a renormalization group treatment supports this picture \cite{Chubukov:2008ve}.

One of the hallmarks of the BCS theory of ordinary superconductors \cite{BCS} was the prediction of an energy gap to single particle excitations. The theory was extended by Mattis and Bardeen \cite{Mattis-Bardeen} to calculate the electrical conductivity as a function of temperature and frequency. They predicted that just below T$_c$, for frequencies below the gap $2\Delta$, the real part of the conductivity has an enhancement from ``coherence factor" effects. These originate from the interference between the transition amplitudes of the electromagnetic field operator $\bm{A}$ between single particle states. This occurs because the superconducting ground state is a \textit{coherent} superposition of single particle states \cite{tinkhambook} which participate in the scattering and absorption of the field.

In general, a coherence peak can arise when the portions of Fermi surface that are coupled by an experimental probe have gaps of the same sign and magnitude.  The first evidence for it was in the NMR relaxation rate of Al in 1958 \cite{Hebel:1959zr}. In the AC conductivity, the first observation of this effect was in Tinkham's far infrared absorption experiments of lead \cite{Palmer:1968ly}. In the case of the iron based superconductors, the coherence or Hebel-Slichter peak has not been observed in NMR experiments \cite{Matano_nmr2008,Nakai_nmr2008}.  Because NMR is a local probe and can couple parts of the FS that differ by large momentum transfer, the absence of the coherence peak has been interpreted as supporting the picture of the sign changing extended s-wave symmetry of the gap function (s$\pm$)\cite{Chubukov:2008ve,Mazin:2008fk,Parish:2008nx}. 

In this Letter we report evidence for the coherence peak in the THz conductivity of an iron pnictide \bao.  Our results point to the existence of an order parameter that is more or less uniform over a FS sheet.  We also find evidence for pair-breaking effects in the form of a residual real part of the conductivity at low temperature and a reduced superfluid density over the expected value. Together these results are strongly consistent with an s$\pm$ symmetry of the order parameter which has opposite sign gaps on different FS sheets.

\begin{figure}
\includegraphics[width=.8\columnwidth]{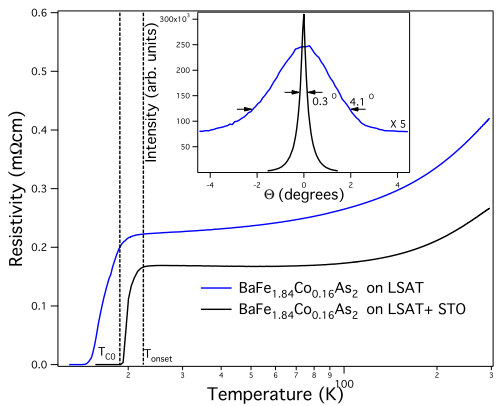}
\caption{Resistivity $\rho(T)$ of \bao\ film as a function of temperature for films grown directly on LSAT (blue (gray)) and grown with the STO template layer (black). Vertical lines indicate the onset (T$_{onset}$) and zero resistance temperatures (T$_{CO}$); T$_c$ is chosen at the midpoint of these 2 values, 20.6 K with a width of 1.8 K.  Inset: X-ray diffraction rocking curve and FWHM for (004) reflection of the same films. The intensity for the LSAT film is multiplied by 5.} \label{resis}
\end{figure}

Thin films of \ba\ were grown by pulsed laser deposition on LSAT substrates with a STO template to allow epitaxial growth.  Extensive details on the film growth can be found elsewhere \cite{Lee:2009qf}.
The novel template engineering results in films of exceptional quality. As shown in the inset to Fig. \ref{resis} the full width at half maximum of the (004) reflection rocking curve of the film on STO/LSAT is as narrow as 0.3$^\circ$, which is the narrowest ever reported for 122 thin films, whereas that of the film on bare LSAT is as broad as 4.1$^\circ$.   The small mosaic spread is comparable to good single crystals.  Residual resistivity values (Fig. \ref{resis}) of films grown on STO/LSAT are almost a factor of 2 smaller than those grown directly on LSAT. For the present film with $x=$0.8, T$_c$ is found to be 20.6 K  with a transition width of 1.8 K.

Terahertz conductivity measurements were performed using a home-built terahertz time domain transmission spectrometer (TTDS) at JHU, which in these films can access the frequency range between 200 GHz and 2 THz. In TTDS, a femtosecond laser pulse is split along two paths and excites a pair of photoconductive Auston-switch antennae deposited on radiation damaged silicon on sapphire. A broadband THz pulse is emitted by one antenna, transmitted through the \ba\ film, and measured at the other antenna.  By varying the length-difference of the two paths, the electric field of the transmitted pulse is mapped out as a function of time. Comparing the Fourier transform of the transmission through \ba\ to that of a reference resolves the full complex transmission coefficient. The transmission is then inverted to obtain the complex conductivity by using the appropriate formula in the thin film approximation. By measuring both the magnitude and phase of the transmission, this inversion can be done directly and does not require Kramers-Kronig transformation.  Reference measurements were performed by ratioing to a nominally identical LSAT/STO layered sample.  Although explicit effects of the three layers in transmission are possible, additional measurements found the influence of the STO layer to be negligible. The temperature dependence was obtained by placing the sample in a flowing vapor cryostat with control of the sample temperature in a range between 1.7-300 K.

\begin{figure}
\includegraphics[width=.75\columnwidth]{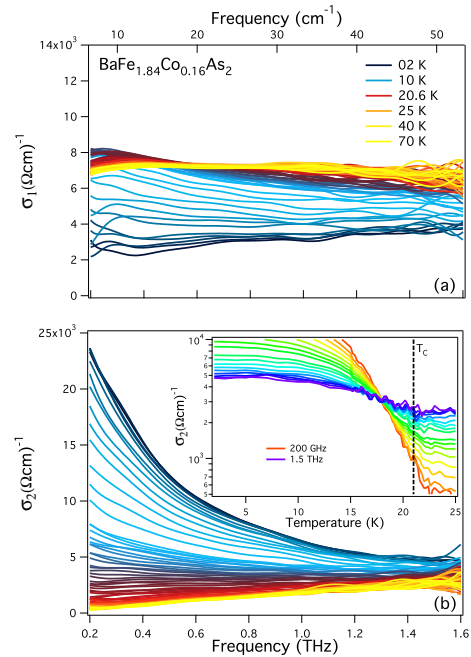}
\caption{(Color Online) (a) Real part of the THz conductivity of \bao. (b). Imaginary part of THz conductivity. Inset. Temperature dependent $\sigma_2$ for several frequencies between 200 GHz and 1.5 THz in 70 GHz steps.} \label{sigmas}
\end{figure}

Fig.\ref{sigmas} shows the measured THz complex conductivity of \bao\ for different temperatures.  At low temperature the imaginary part  exhibits the characteristic 1$/\omega$ dependence expected for a superconductor well below T$_c$.  The real part of the conductivity shows a strong decrease over most of the frequency range, which is reminiscent of a standard s-wave superconductor.  However, at the lowest temperature (2 K), the conductivity is not fully suppressed.  As we will be discussed below, we interpret this as an indication of pair-breaking effects. Significantly, at low frequencies and slightly below T$_c$, the real conductivity initially shows a small enhancement before decreasing at low temperature.   We believe this enhancement is due to the coherence peak effect.

\begin{figure}
\includegraphics[width=.9\columnwidth]{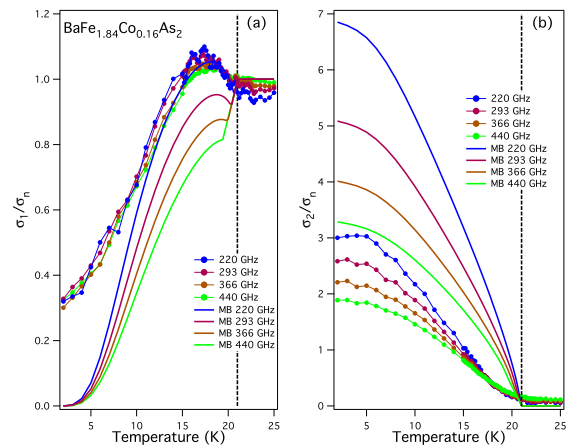}
\caption{(Color Online) (a) Real and (b) imaginary parts of the THz conductivity as a function of temperature for different frequencies. Solid lines are calculations using the standard Mattis-Bardeen form.} \label{mb}
\end{figure}

In Fig. \ref{mb} we show the temperature dependence of the THz conductivity for several frequencies below 2$\Delta$. One can clearly see that just below T$_c$ the conductivity increases above the normal state value (plots were normalized to the 21 K conductivity). Solid lines in Fig. \ref{mb} are fits using the standard Mattis-Bardeen form with a single gap, with $\Delta$= 3 meV, however fits of similar quality can be obtained using 2 gaps. Although the theoretical curve qualitatively resembles the data near T$_c$, at low temperatures the fit is not satisfactory.  The theoretical form shows an exponential suppression of the real part of the conductivity at low temperature while the data show a finite residual value of approximately 30\%.  Although similar residual conductivities have been observed \cite{Gorshunov:2010} and interpreted in terms of nodes in the order parameter \cite{Reid:2010oq,Fischer:2010}, as we argue below we believe that the residual conductivity derives from pair breaking effects.

It is also interesting to note the very narrow temperature region over which superconducting fluctuations are exhibited.  As shown in the inset of Fig. \ref{sigmas}(b), $\sigma_2$ becomes enhanced in region only approximately 2 K above T$_c$.  There is also no sign of an above T$_c$ `dissipation peak' arising from superconducting fluctuations, cf. Fig. \ref{mb}(a).  The narrow range of SC fluctuations may be a reflection of the 3D character of superconductivity in these materials, and are very different from the broad fluctuation range seen via TTDS in the cuprate family of high-T$_c$ superconductors and amorphous thin films \cite{Corson1999,Luke,Crane07a}.

The observation of the coherence peak in the AC conductivity of \ba\ is an important clue to the symmetry of the superconducting parameter in this family of materials.  A coherence peak is usually taken as evidence of a uniform gap function, although more precisely what is required is that portions of the Brillouin zone coupled by the experimental probe have gaps of the same sign and similar magnitude.  Because in general optical measurements probe only zero-momentum excitations around the Fermi surface, in the case of the predicted s$\pm$ symmetry in the pnictides superconductors, one expects the conductivity to only detect a single sign of the order parameter, since the different sheets are separated by large momentum transfer.   Since there is no sign change in the order parameter within a single sheet, we expect a coherence peak that qualitatively resembles that from a single-uniform-gap superconductor.

It is interesting to compare our observation of a coherence peak to the expected one in NMR relaxation rate i.e. the Hebel-Slichter peak. A number of groups \cite{Ahilan:2008cr,Ning-122-nmr} have reported a lack of a coherence peak in the dissipation below T$_c$. We can interpret these results as follows: as NMR is a local probe that excites a single-species nuclear magnetic resonance, it is sensitive to superposition of different momenta as well as to the momentum-dependent structure factor of the nuclei. Thus, NMR relaxation rate is affected by both inter- and intraband processes weighted by the appropriate nuclear structure factors. The absence of the coherence peak is natural when gap-sign-changing interband processes dominates the response, as predicted by \cite{Mazin:2008fk,Chubukov:2008ve}.   It remains to be explained however, how there could be such a minor contribution to intraband processes from all nuclei \cite{Parish:2008nx}.

The existence of an s$\pm$ order parameter however is superficially at odds with the residual real conductivity at low temperature. Although such residual conductivity may be consistent with nodes in the order parameter, we feel that overall the data are more consistent with pair-breaking effects.   In other cases where nodal excitations have been conclusively identified, like in the $d$-wave high-T$_c$ superconductors, the intrinsic scattering rates are very small, with the most extreme case being YBCO ($<$ 2 GHz) \cite{Hosseini99a}.  That is not the case here, as the width of any low temperature, low frequency peak is larger than our spectral range (1.8 THz).   In contrast, one may expect that pair-breaking effects from impurity scattering are substantial in a superconductor with positive and negative contribution to its order parameter \cite{Mazin:2008fk,Chubukov:2008ve}. We expect scattering in these materials because of the effect of Co doping, and in particular for this films, due to the presence of columnar defects along the c-axis \cite{Lee:2009qf}.

Even more significant evidence for substantantial pair-breaking comes from the imaginary part of the conductivity (Fig. \ref{mb}(b)) (which is proportional to the superfluid density), since it is greatly suppressed as compared to the BCS result.  In its essence, the Matthis-Bardeen formalism and its dependence on the size of the superconducting gap, is a statement about the conservation of spectral weight in the superconductor.  In order to estimate the effects of pair breaking, we calculate the spectral weight missing from the superfluid condensate by subtracting the measured $\sigma_2$ from the calculated MB $\sigma_2^{MB}(T=0)$ as $\hbar\omega[\sigma_2^{MB}(T=0)-\sigma_2^{measured}(T=0)]$. If pair-breaking is the major contributor to the reduced superfluid density, this difference should be approximately equal to $\frac{\pi}{2}\sigma_1(T=0)\times2\Delta$ (see Fig. \ref{mb}a). We find that the residual $\sigma_1$ accounts for a substantial 75\% of the missing superfluid density from the MB expectation. We expect this estimation to give a lower bound on the effects of pair-breaking.  We find therefore that observation of a residual $\sigma_1$ is consistent with an s$\pm$ pairing symmetry of the gap without nodes.  Additional support for our interpretation comes from a comparison with far infrared conductivity studies of conventional $s$-wave superconductors doped with magnetic impurities.  For an extended s-wave superconductor, non-magnetic impurities should act the same as magnetic impurities in an s-wave superconductor.  The present data bear a strong resemblance to the far infrared conductivity of Pb films doped with Mn impurities \cite{Dick:1969fk,Woolf:1965fk}, where a residual value in the zero frequency limit in $\sigma_1$ was found \cite{note1}, but it is in stark contrast with the prediction of Abrikosov and Gorkov \cite{AG} where one expects $\sigma_1(\omega\rightarrow0)=0$.

\begin{figure}
\includegraphics[width=.9\columnwidth]{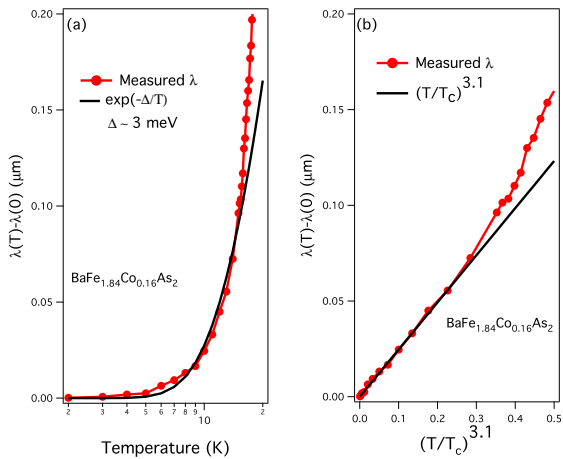}
\caption{(Color Online) London penetration depth of \bao\ film. (a) Comparison of measured $\lambda$ with exponential function of temperature, a gap value of $\Delta$ can be extracted with a fit below 5 K as shown. (b) Data plotted against (T/T$_c$)$^{3.1}$.} \label{pend}
\end{figure}

The temperature dependence of the penetration depth supports this scenario as well.  Fig. \ref{pend} shows the penetration depth $\lambda(T)$ extracted from the imaginary part of the conductivity at 220 GHz.  At low temperature $\lambda$ behaves almost exponentially with a thermally activated gap $\Delta\approx 3$ meV. However, note that the data follow a power law behavior with an exponent of 3.1 (Fig. \ref{pend}(b)).   Although power laws in the penetration depth are typically taken as indicative of the presence of nodes in an order parameter, such a large exponent is generally not consistent with dirty nodal behavior \cite{Hirschfeld:1993}. In the present case we think that the large exponents are indicative of pair breaking excitations. This view is supported by a recent study of the penetration depth measured as a function of Pb radiation induced disorder \cite{Kim:2010vn} in single crystals of \ba. The authors found that the exponent decreased with increasing disorder and that cleaner samples have a larger exponent and was consistent with pair breaking effects. Based on a model of s$\pm$ symmetry of the gap without nodes, they were able to explain the trend with disorder. Although our large exponent power-law of more than 3 is in contrast to microwave and far infrared measurements in single crystals \cite{Gordon:2009bh} and thin films \cite{Fischer:2010}, where a smaller exponent closer to 2 has been observed, it is consistent with the excellent quality of our films. In addition, calculations \cite{Dolgov-2009,Vorontsov:2009fk} of the microwave conductivity and penetration depth that take into account the pair-breaking scattering due to impurities in a multiband s$\pm$ superconductor, reproduce both our observations of finite absorption below the gap (c.f. Fig.\ref{mb}(a)) and a power law dependence of the penetration depth (c.f. Fig. \ref{pend}(b)) at low temperatures. This calculation places the impurity scattering in \ba\ between the Born and unitary limits.

We have reported evidence that strongly support the extended s-wave scenario for the symmetry of the superconducting gap in \bao. In particular the presence of the coherence peak in the THz conductivity, the power law behavior of the penetration depth $\lambda$ with a high exponent and presence of pair-breaking scattering are consistent with an s$\pm$ symmetry of the order parameter which does not contain nodes.   We hope that our work stimulates further investigation into the effects of pair-breaking in these compounds.   In particular, detailed calculations of their effects on the conductivity in the superconducting state would be very useful.

This work was supported under the auspices of the Institute for Quantum Matter under grant DOE DE-FG02-08ER46544 at JHU. Work at UW was supported by DOE DE-FG02-06ER463 and work at NHMFL supported by NSF DMR-0084173, by the State of Florida and by AFOSR FA9550-06-10474. We would like to thank A. Bernevig, G. Boyd, R. Gordon, K. Moler, J. Murray, V. Stanev, Z. Te\v{s}anovi\'c and D. Wu for useful discussions.

\bibliography{Ba122}
\end{document}